\documentclass[12pt, letterpaper]{amsart}
\usepackage[left=.8in,right=.8in,bottom=0.5in,top=0.8in]{geometry}
\usepackage{amsfonts}
\usepackage{amsfonts,amsbsy,amssymb,amsmath,graphicx,amsthm, bm} 
\usepackage{graphicx}
\usepackage[font=small,labelfont=bf]{caption}
\usepackage{epstopdf}
\usepackage[pdfpagelabels,hyperindex]{hyperref}
\usepackage{xcolor}

\usepackage{float}
\usepackage{pgfplots}
\usepackage{listings}
\usepackage{longtable}
\usepackage{mathrsfs}
\usepackage{mathtools}
\usepackage{comment}

\usepackage{subcaption}
\usepackage{lipsum}

\usepackage{epsfig}
\usepackage{dsfont}
\usepackage{natbib}
\usepackage{mathrsfs}
\usepackage{geometry}

\theoremstyle{definition}

\theoremstyle{remark}

\definecolor{energy}{RGB}{114,0,172}
\definecolor{freq}{RGB}{45,177,93}
\definecolor{spin}{RGB}{251,0,29}
\definecolor{signal}{RGB}{203,23,206}
\definecolor{circle}{RGB}{217,86,16}
\definecolor{average}{RGB}{203,23,206}

\colorlet{shadecolor}{gray!20}
\pgfplotsset{compat=1.9}

\usepgflibrary{fpu}

\def\1{1\kern-.20em {\rm l}}
\usepackage{enumitem}
\usepackage{lscape}
\usepackage{booktabs}

\numberwithin{equation}{section}

\author[Mohamed Chaouch]{Mohamed Chaouch}
\address{(M. Chaouch) Statistics Program, Department of Mathematics and Statistics, College of Arts and Sciences, Qatar University, Doha, Qatar.}
\email{mchaouch@qu.edu.qa}
\author[Thanasis Stengos]{Thanasis Stengos}
\address{(T. Stengos) Department of Economics and Finance, University of Guelph, Canada}
\email{tstengos@uoguelph.ca}
\keywords{}
\urladdr{}
\title[Modeling the Happiness-Sustainability Nexus]{Modeling the Happiness-Sustainability Nexus via Graphical Lasso and Quantile-on-Quantile Regression}

\begin{document}
\maketitle

\begin{abstract}

This paper investigates the nexus between subjective well-being and sustainability, proxied by the Sustainable Development Goals (SDG) Index, using cross-country data from 126 nations in 2022. While prior research has highlighted a positive association between happiness and sustainable development, existing approaches largely rely on linear regressions or correlation-based measures that mask distributional heterogeneity, multicollinearity, and potential nonlinear dependence. To address these 
limitations, we employ a two methodological framework combining Graphical Lasso, and Quantile-on-Quantile Regression (QQR). The Graphical Lasso identifies a direct conditional link between happiness and sustainability after controlling for governance, income,  and life expectancy, with a partial correlation of about 0.21. On the other hand, QQR reveals heterogeneous effects across the joint distribution: sustainability gains are 
positively associated with happiness for low-happiness but high-sustainability countries, negatively associated in high-happiness but low-sustainability contexts, and essentially neutral elsewhere. These findings suggest that the happiness–sustainability link is modest, asymmetric, and context-dependent, underscoring the importance of moving beyond mean-based regressions. From a policy perspective, our results highlight that institutional quality, income, and demographic factors remain the dominant 
drivers of both happiness and sustainability, while the interplay between the two dimensions is most pronounced in distributional extremes.

\end{abstract}

\bigskip
\noindent {JEL classification:} E20, I31, Q56

\noindent {Keywords}: Graphical Lasso, Quantile-on-Quantile Regression,  Happiness, SDG Index.

\section{Introduction}
\subsection{Problem statement} How closely do countries’ achievements in sustainable development translate into greater subjective well-being? This question has gained prominence as policymakers and scholars move beyond income-centric metrics toward multidimensional assessments of progress. The World Happiness Report (WHR) synthesizes Gallup World Poll measures of life evaluation and repeatedly documents strong associations between well-being and socioeconomic and institutional factors, while the Sustainable Development Report (SDR) provides the SDG Index, a composite measure of performance across the 17 Sustainable Development Goals. Yet the form of the relationship between country-level happiness and the SDG Index-linear or nonlinear, symmetric or tail-driven, direct or mediated by covariates-remains imperfectly understood.\\
Recent evidence suggests a positive global association between progress on the SDGs and subjective well-being, with heterogeneity across specific goals and regions. For example, \cite{denevesachs2020} report strong positive links between most SDGs and life satisfaction, alongside trade-offs for SDG12 (Responsible Consumption and Production) and SDG13 (Climate Action), underscoring that the happiness–sustainability relationship is unlikely to be well captured by a single linear model. These patterns motivate methods that can distinguish direct from mediated dependence, accommodate nonlinearities, and probe tail co-movements. \\
Standard cross-country regressions or simple correlations are ill-suited to these tasks. First, they deliver average effects and therefore mask distributional heterogeneity across low- and high-performing countries. Second, they cannot easily separate direct conditional links (e.g., between happiness and the SDG Index) from indirect links operating through income,   health, or governance. Third, classic Gaussian correlation is insensitive to asymmetric and tail dependence that are central to policy (e.g., whether very sustainable countries disproportionately occupy the top of the happiness distribution).\\
To address these limitations, we propose a two, complementary framework. We first use the Graphical Lasso to map direct conditional associations in a sparse network of happiness, the SDG Index, and key covariates; next, we deploy Quantile-on-Quantile Regression to expose distributional heterogeneity in the happiness–SDG link. This design ensures that each method answers a different question-who is directly linked, where in the distributions the link is strongest, and what the dependence looks like in full-thereby avoiding redundancy while delivering a unified empirical narrative.\\
Our empirical application uses the most recent WHR and SDR country data to characterize the cross-sectional dependence between happiness and the SDG Index conditional on socioeconomic and governance covariates. We show that the direct edge between happiness and sustainability persists after conditioning, that the association is markedly stronger in specific quantile regimes, and that the conditional dependence displays meaningful tail features-insights that materially enrich policy interpretation beyond mean effects.

\subsection{Organization of the paper}
The remainder of the paper is organized as follows. Section~\ref{sec2} reviews the literature on the happiness–sustainability nexus. Section~\ref{sec3} describes the
data and exploratory analysis. Section~\ref{meth} presents the methodological framework, covering Generalized Additive Models, Graphical Lasso, and QQR. Section~\ref{res} reports the empirical results, and Section~\ref{conc} concludes with key findings and policy
implications. Finally, the definition and source of data are given in the Appendix.

\section{Literature Review}\label{sec2}
The received literature on happiness and sustainability is still relatively thin and fragmented, with much of it dominated by conceptual or normative arguments. Only in recent years have scholars begun to empirically interrogate the link between subjective well-being and sustainable development at the cross-country level.
\subsection{Conceptualizing the Happiness-Sustainability Nexus}
Several contributions approach the happiness–sustainability nexus from a theoretical or philosophical perspective. In a foundational essay, \cite{bartolini2014building} challenges the common perception that happiness and sustainability are inherently incompatible. He identifies mechanisms-such as high discount rates, lack of trust in collective action, and the erosion of confidence in political institutions-that jointly produce a bias toward unsustainable consumption, even when such practices undermine long-run well-being. The core argument is that individual time preferences, institutional weaknesses, and the concentration of economic and political power lead to unsustainable equilibria that fail to secure happiness across generations.\\
In a similarly normative register, \cite{brulde2015well} examines whether the pursuit of a more sustainable lifestyle necessarily reduces well-being. He notes that in the short run, stricter sustainability targets (e.g., drastic cuts in carbon emissions per capita) imply substantial changes in consumption habits and mobility patterns, likely lowering life evaluations. Yet drawing on the adaptation literature \citep{frederick199916}, he argues that individuals may, over time, adjust to these new standards of living, restoring happiness levels. Sustainability, from this angle, need not be seen as an enduring sacrifice of well-being but rather as a transition requiring adaptation.\\
From a socio-economic angle, \cite{barrington2016sustainability} calls for a cultural transformation in which capitalist structures are replaced by a “zero marginal cost society.” In such a system, technology and automation would reduce the cost of producing goods, increasing the role of public goods and collective assets. This scenario is expected to generate more pro-social and creative behaviors, which are shown to foster subjective well-being. Following the reasoning of \cite{akerlof2015phishing}, reducing the manipulation of preferences by advertising and private profit maximization would align human behavior more closely with both sustainability and happiness. These conceptual accounts highlight the complex interplay between institutions, consumption patterns, and cultural norms in shaping the joint pursuit of sustainability and well-being.
\subsection{Empirical Evidence on Happiness and Sustainability}
Empirical work on the happiness–sustainability relationship is more recent and comparatively sparse. An early contribution by \cite{zidanvsek2007} documented positive correlations between various measures of happiness (national happiness levels, inequality-adjusted happiness, happy life years) and environmental performance (measured by the Environmental Sustainability Index and the Environmental Performance Index). He concluded that happier nations are not only more environmentally conscious but also more energy efficient, suggesting that sustainability and well-being need not involve intergenerational trade-offs. More recent correlation-based studies, such as \cite{lakocai2023sustainable}, also report positive associations between ecological footprint and well-being, though the ecological footprint is widely seen as an incomplete proxy for sustainability, being closely tied to economic development levels.\\
Other contributions have attempted to formalize the analysis using econometric frameworks. \cite{dietz2009environmentally} employed a Stochastic Impacts by Regression on Population, Affluence, and Technology (STIRPAT) modeling framework to explore how sustainability factors shape well-being, with affluence, education, and life expectancy as controls. While innovative, the study’s reliance on life expectancy as the main proxy for well-being limits its interpretability. Nevertheless, it demonstrated the feasibility of incorporating sustainability variables into systematic models of national happiness.\\
More directly relevant is the work of \cite{helliwell2016}, who examined the contribution of sustainable development to happiness using the SDG Index alongside Gallup World Poll life evaluation data. His results suggested that the SDG Index alone explains a substantial share of cross-country variation in happiness. Follow-up studies such as \cite{sameer2021happier} confirmed a positive link through OLS estimates, though more sophisticated methods (2SLS and GMM) yielded weaker or insignificant coefficients, raising questions about causality and potential endogeneity.\\
Disaggregating the SDGs, \cite{aksoy2020evaluation} used structural equation modeling to assess how economic, social, and environmental sustainability dimensions relate to happiness, proxied by the Happy Planet Index. They found both social and environmental sustainability to be positively correlated with well-being, strengthening the case for multidimensional assessments. Similarly, \cite{lin2024economic} focused on OECD countries and employed data envelopment analysis (DEA) to link the OECD Better Life Index to sustainable development. Their results showed that several countries achieved both high well-being efficiency and strong sustainability performance, although the method could not disentangle causal pathways or conditional dependencies.

\subsection{Identified Research Gap}
Despite this growing body of evidence, several gaps remain. First, most empirical studies rely on average correlations or mean-based regression models, which cannot distinguish whether the happiness–sustainability link is direct or merely reflects shared covariates such as income, education, or governance. Second, little attention has been paid to distributional heterogeneity-for instance, whether sustainability matters more for low-happiness countries than for high-happiness ones. Third, the potential for nonlinear and asymmetric dependence, particularly in the tails of the distributions, remains unexplored despite evidence of trade-offs for specific SDGs such as climate and consumption. Addressing these limitations requires methods capable of identifying direct conditional links, uncovering heterogeneity across quantiles, and characterizing nonlinear and tail dependence. Our study contributes to filling these gaps by applying complementary empirical strategies to systematically reassess the happiness–sustainability nexus.

\section{Data Description and Exploration}\label{sec3}

The empirical analysis relies on a cross-country dataset that combines measures of subjective well-being, sustainability performance, governance quality, and socioeconomic development. The dependent variable is Happiness, measured by the Cantril Ladder of Life Evaluations from the Gallup World Poll, which reflects individuals’ self-assessed life satisfaction on a scale from 0 (worst possible life) to 10 (best possible life). The key explanatory variable of interest is the Sustainable Development Goals (SDG) Index, compiled by the UN Sustainable Development Solutions Network (SDSN), which summarizes each country’s progress toward achieving the 17 SDGs on a 0–100 scale, with higher values indicating stronger sustainability performance.\\
To account for confounding factors, the dataset also includes a range of governance indicators from the Worldwide Governance Indicators (WGI): Government Effectiveness (GE), Regulatory Quality (RQ), Rule of Law (RL), Voice and Accountability (VA), and Control of Corruption (CC). These variables capture institutional capacity and democratic quality, which are known to influence both happiness and sustainable development outcomes. In addition, we include socioeconomic and demographic variables such as Life Expectancy (LE), Domestic Credit to Private Sector (DCPS, \% of GDP), and the Index of Economic Freedom (IEF), reflecting health, financial development, and market liberalization, respectively.\\
The sample covers 126 countries with data available for the year 2022, ensuring global coverage across high-, middle-, and low-income economies. After excluding cases with missing values, the final dataset includes 14 numeric variables and 78 complete observations, which form the basis for the Graphical Lasso, and Quantile-on-Quantile Regression analyses. 

\subsection{Correlation analysis and variables' distribution}
A preliminary inspection using scatterplot matrices reveals substantial heterogeneity across countries and strong correlations among governance indicators, underscoring the importance of applying methods that can handle multicollinearity and uncover nonlinear dependence structures.

\begin{figure}[h!]
    \centering
    \includegraphics[width=0.95\textwidth]{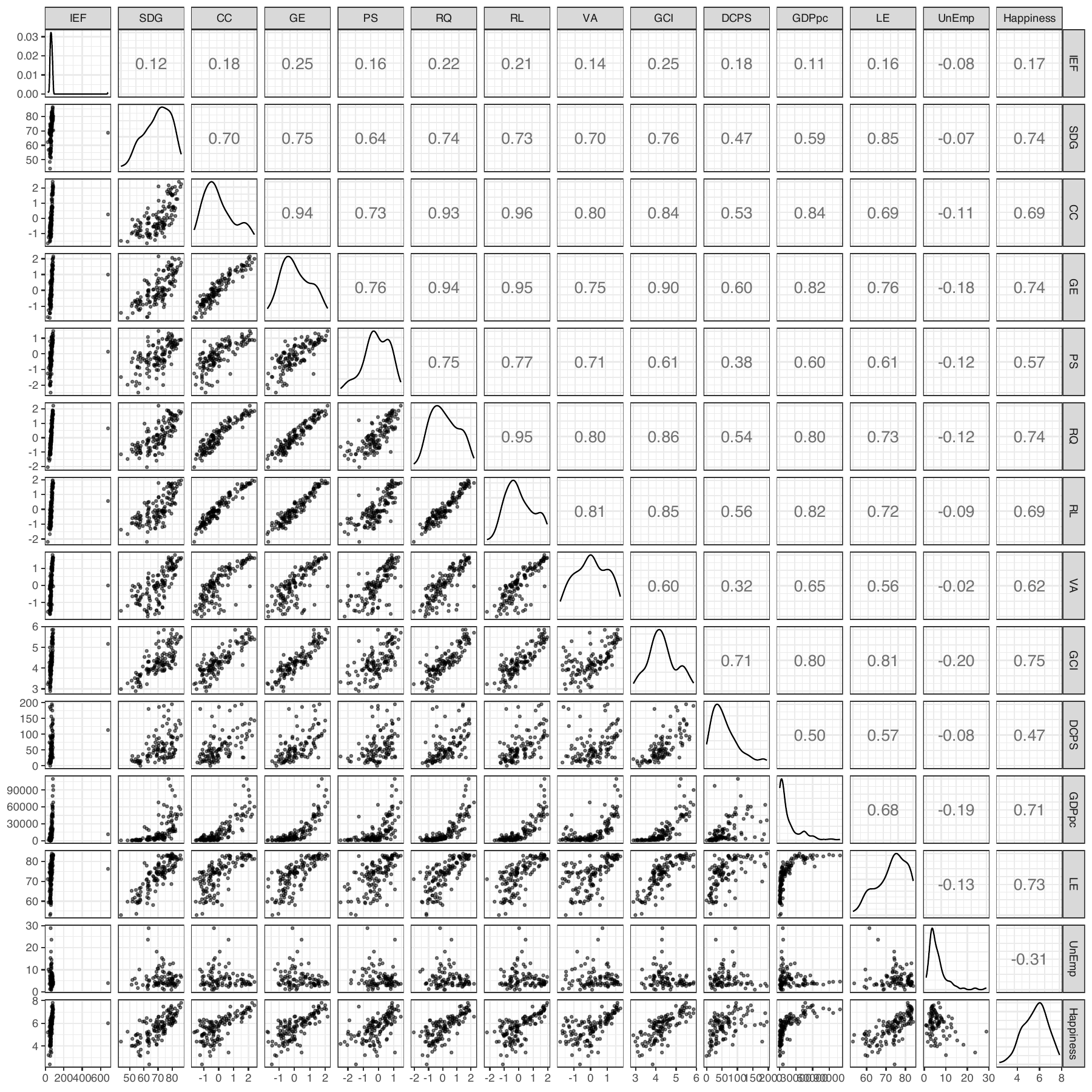}
    \caption{Scatterplot matrix of happiness, SDG Index, and related predictors. Lower panels show scatterplots, upper panels report correlations, and diagonal panels display univariate densities.}
    \label{fig:scatterplot_matrix}
\end{figure}

The scatterplot matrix in Figure~\ref{fig:scatterplot_matrix} provides an overview of the pairwise relationships between happiness, sustainability (SDG Index), institutional quality measures, and socioeconomic indicators. The diagonal panels display the univariate distributions, which show that most variables are continuous and reasonably well spread, although some (such as unemployment and inequality) exhibit skewness. The upper panels report Pearson correlation coefficients, while the lower panels display 
scatterplots that reveal mostly linear associations with varying dispersion. Several governance indicators (control of corruption, government effectiveness, regulatory quality, rule of law, and voice and accountability) are highly correlated with one another, with coefficients frequently exceeding 0.9, indicating a strong presence 
of multicollinearity within the institutional block. Both happiness and the SDG Index are strongly and positively associated with life expectancy, income, and governance measures, with correlations generally in the range of 0.6 to 0.8. These findings suggest that the observed happiness–sustainability nexus is largely explained by shared institutional and socioeconomic drivers.  \\
At the same time, other socioeconomic variables display more nuanced relationships. Domestic credit to the private sector and the Index of Economic Freedom are positively correlated with both happiness and sustainability, whereas unemployment correlates negatively.  Taken together, the scatterplot matrix underscores two key insights. First, there is a clear positive clustering of happiness, sustainability, governance, and socioeconomic development, reflecting their shared underpinnings. Second, the strong collinearity among 
governance indicators highlights the need for statistical methods that can address overlapping predictors, motivating the use of approaches such as Graphical Lasso, and Quantile-on-Quantile Regression in the subsequent analysis.

\subsection{Clustering Analysis of Countries and extraction of main features}

To explore whether countries can be meaningfully grouped based on their joint performance in happiness, sustainability, and related socioeconomic and institutional factors, we applied an unsupervised learning framework combining k-means and hierarchical clustering. Prior to clustering, all numeric features were standardized to zero mean and unit variance in order to ensure comparability across variables measured on different scales. \\
The optimal number of clusters $k$ was selected using a combination of the elbow criterion and the average silhouette width. The elbow plot examined the total within-cluster sum of squares as a function of $k$, while the silhouette statistic measured how well each country 
fits within its assigned cluster relative to other clusters. Both diagnostics suggested that a two-cluster solution offers the most parsimonious and interpretable representation. For 
robustness, we also compared k-means with hierarchical clustering using Ward’s method, which yielded highly consistent partitions. To aid visualization, we reduced the high-dimensional 
feature space to two principal components (PCA) and projected the cluster assignments onto this map.

\begin{figure}[H]
    \centering
    \begin{minipage}{0.48\textwidth}
        \centering
        \includegraphics[width=\linewidth, height = 7cm]{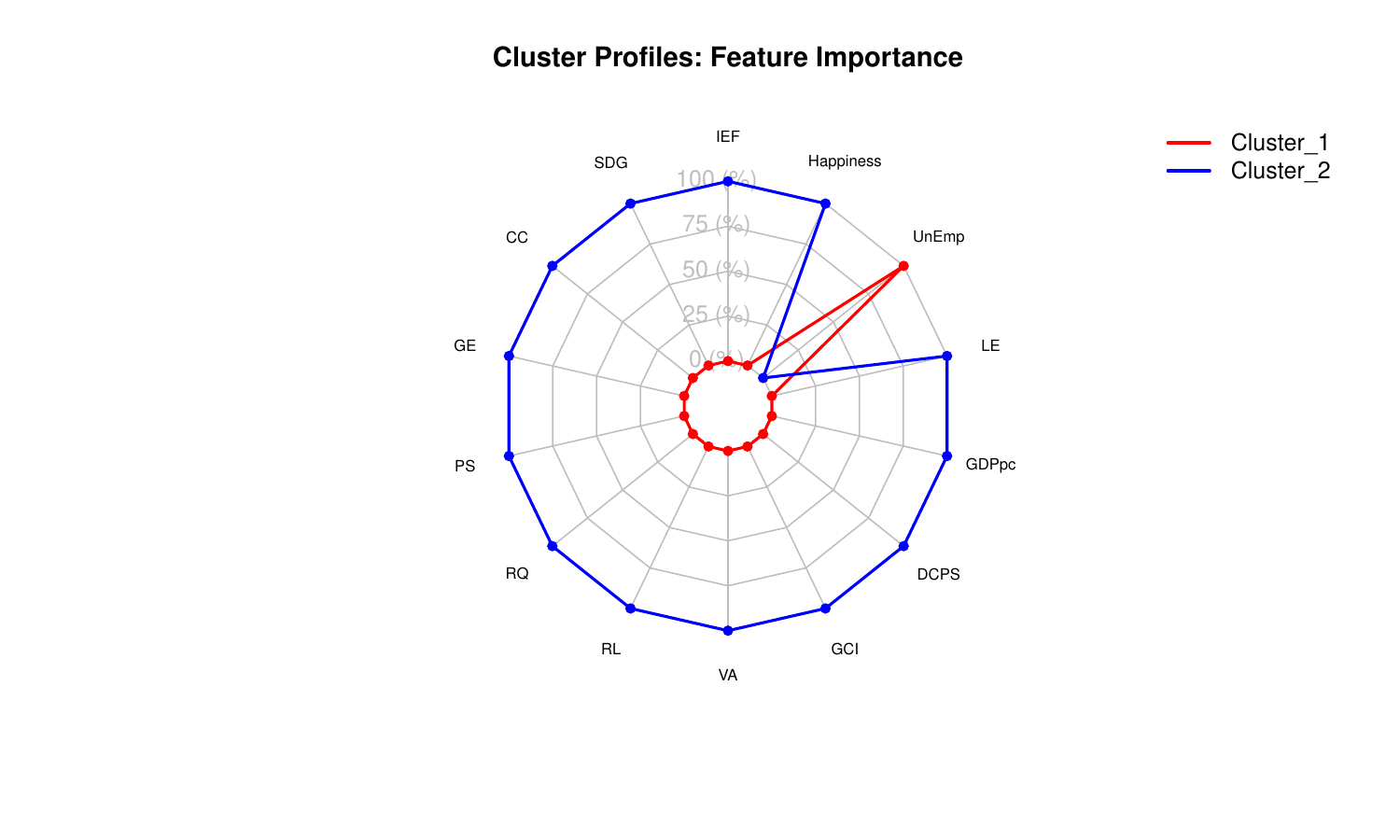}
        \caption*{(a) Cluster profiles across features}
    \end{minipage}\hfill
    \begin{minipage}{0.48\textwidth}
        \centering
        \includegraphics[width=\linewidth, height = 7cm]{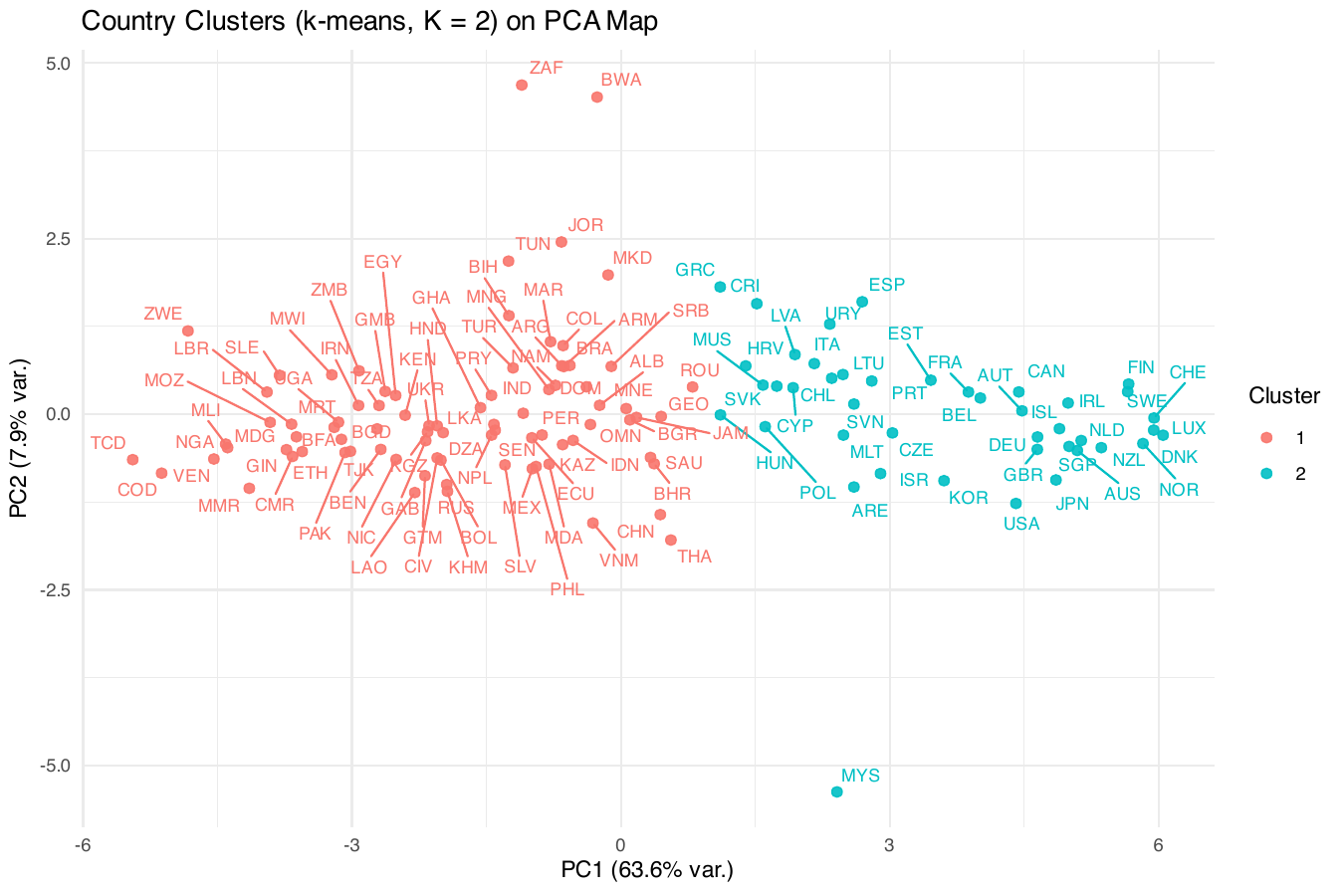}
        \caption*{(b) PCA map of country clusters}
    \end{minipage}
    \caption{Clustering results: (a) radar/heatmap profiles of the features 
    characterizing each cluster and (b) two-dimensional PCA projection showing 
    the spatial separation of countries.}
    \label{fig:clusters}
\end{figure}

The clustering analysis reveals two clearly distinct groups of countries.  Figure~\ref{fig:clusters}(a) shows that Cluster~1 is characterized by  systematically weaker performance across almost all dimensions, with particularly low scores in happiness, sustainability (SDG Index), income (GDP per capita), life expectancy, 
and governance indicators, while displaying markedly higher unemployment rates. By contrast, Cluster~2 exhibits consistently stronger outcomes, with higher happiness levels, stronger SDG performance, better institutional quality (control of corruption, government effectiveness, rule of law, regulatory quality, and voice and accountability), and more favorable economic indicators such as credit provision and competitiveness. These contrasting feature profiles highlight unemployment and institutional quality as the main dimensions separating the clusters. Complementary evidence from the PCA map in Figure~\ref{fig:clusters}(b) confirms that the two clusters are not only statistically well separated but also geographically diverse: countries in Cluster~2 are concentrated in high-income and OECD regions, whereas Cluster~1 includes lower- and middle-income economies with weaker governance structures. Taken together, the clustering results demonstrate that countries differ markedly in their positioning along the happiness–sustainability nexus, with prosperous, well-governed nations achieving a mutually reinforcing balance between well-being and sustainability, while weaker economies face the dual challenge of low well-being and poor sustainability outcomes.

\subsection{Multicollinearity Diagnostics}

Table~\ref{tab:vif_results} presents the variance inflation factors (VIFs) for all predictors included in the OLS model. The conventional rule of thumb is that VIF values above 10 indicate severe multicollinearity problems. As shown in the table,  {Control of Corruption (CC)},  {Government Effectiveness (GE)},  {Regulatory Quality (RQ)},  {Rule of Law (RL)}, and  {Global Competitiveness Index (GCI)} all exceed this threshold, with values ranging from about 11 to nearly 28. These predictors therefore suffer from substantial redundancy and are highly correlated with other covariates in the model. In contrast, variables such as 
 {Political Stability (PS)},  {Domestic Credit to Private Sector (DCPS)}, 
 {GDP per capita},  {Life Expectancy (LE)}, and  {Unemployment (UnEmp)} remain below the critical cutoff, indicating more acceptable levels of collinearity. The presence of such high VIF values among institutional quality indicators suggests that OLS estimates are likely unstable and inflated, providing further justification for the adoption of more flexible and robust approaches such as Generalized Additive Models (GAMs), which are less sensitive to multicollinearity.  

\begin{table}[H]
    \centering
    \caption{Variance Inflation Factors (VIFs) for OLS predictors}
    \label{tab:vif_results}
    \resizebox{\textwidth}{!}{%
    \begin{tabular}{lccccccccccccc}
        \hline
        IEF & SDG & CC & GE & PS & RQ & RL & VA & GCI & DCPS & GDPpc & LE & UnEmp \\
        \hline
        1.22 & 5.18 &  {15.43} &  {15.47} & 3.50 &  {16.98} &  {27.89} & 4.20 &  {10.87} & 2.54 & 4.33 & 5.33 & 1.16 \\
        \hline
    \end{tabular}}
\end{table}

\subsection{Identifying Influential Countries with Cook’s Distance}
\begin{figure}[h!]
    \centering
    \includegraphics[width=0.6\textwidth]{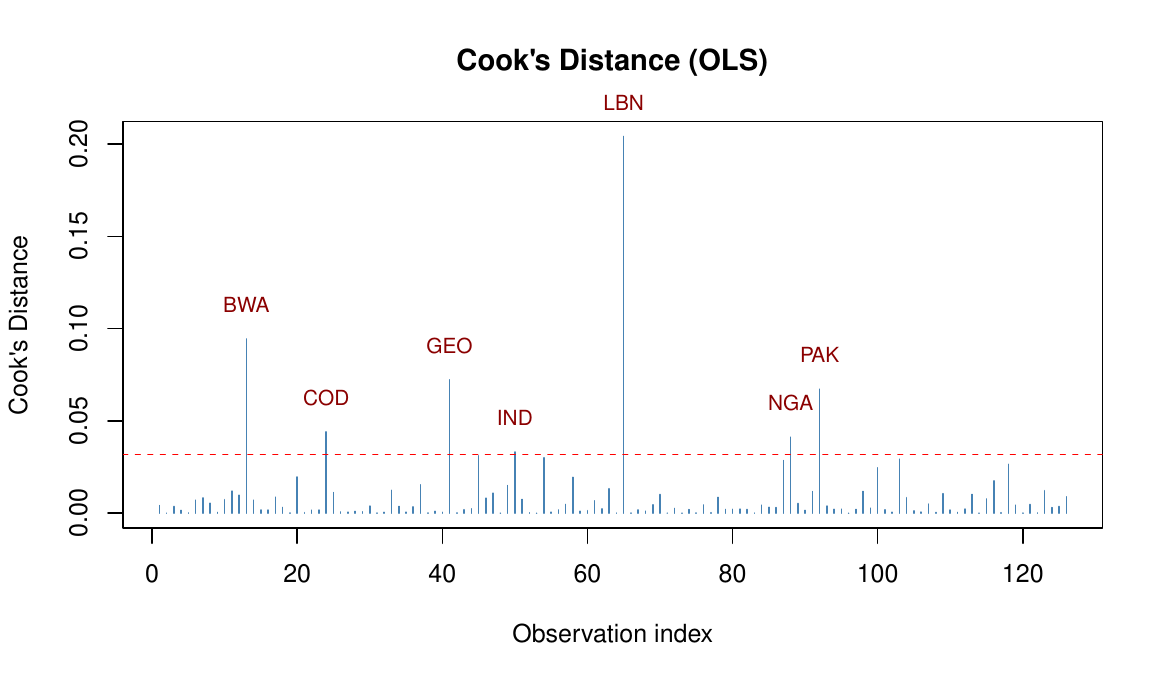}
    \caption{Cook's Distance from the OLS regression of happiness on institutional and socioeconomic predictors. 
    Countries above the dashed red line ($4/n$) are labeled as influential observations.}
    \label{fig:cooks}
\end{figure}

To assess whether a few countries exert disproportionate influence on the regression results, Figure~\ref{fig:cooks} plots Cook's Distance values from the OLS model of happiness. The red dashed line indicates the conventional threshold of $4/n$, where $n$ is the sample size. Most countries lie below this cutoff, but several---including Botswana (BWA), Congo (COD), Georgia (GEO), India (IND), Lebanon (LBN), Nigeria (NGA), and Pakistan (PAK)---exceed it and are labeled in the figure. 

These observations exert greater leverage on the estimated coefficients, suggesting that 
OLS-based results should be interpreted with caution and validated using robust methods. 
Therefore, in the remainder of this paper we employ a Generalized Additive Model (GAM) 
framework in place of ordinary least squares. GAMs offer several advantages in this context: they relax the restrictive linearity assumption by allowing smooth, data-driven relationships between predictors and outcomes, reduce the risk that apparent outliers are merely artifacts of model misspecification, and provide a more robust fit in the presence of heterogeneous functional forms across countries. By capturing potential nonlinearities in the happiness–sustainability relationship, the GAM approach mitigates the undue influence of individual countries and yields results that are more reliable for policy interpretation.

\section{Methodology}\label{meth}

\subsection{Generalized Additive Model} 
Generalized Additive Models, introduced by \citet{hastie1986generalized} and 
further developed in \citet{wood2017generalized}, provide a flexible extension of 
generalized linear models by allowing for nonlinear relationships between predictors 
and the response. Formally, let $Y$ denote a response variable with mean 
$\mu = \mathbb{E}[Y|X]$ linked to predictors $X = (X_1,\ldots,X_p)$ through a monotone 
link function $g(\cdot)$. A GAM has the form
\[
g(\mu) \;=\; \beta_0 + \sum_{j=1}^p f_j(X_j),
\]
where $\beta_0$ is an intercept and each $f_j$ is a smooth, potentially nonlinear 
function estimated from the data. The functions $f_j$ are typically represented 
using basis expansions (e.g., cubic regression splines) 
and estimated by penalized likelihood, with smoothing parameters selected to balance 
model fit against smoothness, often via restricted maximum likelihood.

In our study, GAMs were employed at several stages to capture nonlinear associations 
between happiness, sustainability (SDG Index), and a wide set of socioeconomic and 
institutional predictors. First, prior to quantile-on-quantile analysis, 
GAMs were used to \emph{partial out} the effects of control variables (e.g., GDP per capita, 
  life expectancy, and governance indicators) from both happiness and SDG. 
This ensured that subsequent dependence analyses focused on the residual relationship 
between happiness and sustainability rather than being confounded by common 
socioeconomic drivers. Second, GAMs served as a robustness check against linear 
specifications: by flexibly modeling predictor effects, they allowed us to test 
whether nonlinearities significantly altered the conditional structure of happiness 
and SDG. Finally, in diagnostic analysis, GAMs helped mitigate the influence of 
outliers and influential observations detected under linear models (e.g., via Cook’s 
distance), yielding more stable estimates of the conditional mean structure.

The decision to rely on GAMs rather than ordinary least squares regression is 
further justified by the exploratory diagnostics reported earlier in Section \ref{sec3}. Variance Inflation 
Factors revealed substantial multicollinearity among predictors, while Cook’s 
distance identified influential country-level observations exerting disproportionate 
leverage on OLS estimates. These issues undermine the reliability of linear regression. 
By contrast, GAMs mitigate multicollinearity through penalized smooth estimation and 
offer robustness against influential points by allowing nonlinear functional forms 
that accommodate heterogeneous patterns across countries. Consequently, all subsequent 
conditional modeling of happiness and sustainability is based on GAMs rather than OLS.

\subsection{Graphical Lasso}
A central question in this study is whether happiness and sustainability (proxied by the SDG Index) share a direct association once the influence of key socioeconomic and institutional variables-such as GDP per capita,   life expectancy, and governance quality-is accounted for. Simple correlations cannot answer this question, as they conflate direct and indirect effects, while conventional regressions focus on mean responses and impose linearity. To isolate direct connections, we adopt a Graphical Lasso framework.

\subsubsection{Graphical Lasso Model}

Graphical models provide a rigorous statistical framework for representing the dependence structure among a set of random variables through a network, where nodes denote variables and edges encode conditional dependencies. In the Gaussian setting, this structure is captured by the \emph{precision matrix}, i.e. the inverse of the covariance matrix. Formally, let
\[
X = (X_1, \ldots, X_p) \sim \mathcal{N}_p(0, \Sigma),
\] 
with covariance matrix $\Sigma$ and precision matrix $\Theta = \Sigma^{-1}$. The key property is that $\Theta_{ij} = 0$ if and only if variables $X_i$ and $X_j$ are conditionally independent given all other variables. Hence, estimating a sparse precision matrix is equivalent to uncovering the underlying conditional independence graph.  \\
The \emph{Graphical Lasso} (GLasso), introduced by \cite{friedman2008sparse}, is a regularized maximum likelihood estimator that provides a computationally efficient approach to this problem. Specifically, it solves the following convex optimization program: 
\[
\widehat{\Theta} \;=\; \arg\max_{\Theta \succ 0} \Big\{ \log \det (\Theta) \;-\; \mathrm{tr}(S\Theta) \;-\; \lambda \|\Theta\|_1 \Big\},
\]
where $S$ is the empirical covariance matrix, $\log \det (\Theta)$ enforces positive definiteness, $\mathrm{tr}(S\Theta)$ corresponds to the Gaussian log-likelihood, and $\|\Theta\|_1 = \sum_{i \neq j} |\Theta_{ij}|$ denotes the $\ell_1$ penalty applied to off-diagonal elements. The tuning parameter $\lambda \geq 0$ controls the degree of sparsity: larger values of $\lambda$ yield sparser precision matrices.  \\
The $\ell_1$ penalty shrinks small entries of $\Theta$ towards zero, producing a sparse estimate $\widehat{\Theta}$ where many off-diagonal elements are exactly zero. This sparsity translates directly into the network representation, where an edge between nodes $i$ and $j$ exists if and only if $\widehat{\Theta}_{ij} \neq 0$. Furthermore, from $\widehat{\Theta}$ one can compute partial correlations as
\[
\rho_{ij \cdot V\setminus \{i,j\}} = -\frac{\widehat{\Theta}_{ij}}{\sqrt{\widehat{\Theta}_{ii}\widehat{\Theta}_{jj}}},
\]
which quantify the direct association between variables $i$ and $j$ after conditioning on all others.  \\
From an implementation perspective, estimation of $\widehat{\Theta}$ can be carried out using block coordinate descent algorithms, as described by \cite{friedman2008sparse}, or via more recent scalable solvers such as QUIC (\cite{hsieh2014quic}). Selection of the tuning parameter $\lambda$ is typically performed through information criteria such as Bayesian Information Criterion (BIC) or Extended Bayesian Information Criterion (EBIC), or via stability-based methods such as Stability Approach to Regularization Selection (StARS) (\cite{liu2010stability}).  \\
Graphical Lasso is particularly valuable in high-dimensional settings where the number of variables is large relative to the sample size, since it regularizes estimation, avoids overfitting, and yields an interpretable sparse graphical representation of the conditional dependence structure. In the context of our application, GLasso enables us to assess whether happiness and sustainability (proxied by the SDG Index) are directly connected once a broad set of socioeconomic and institutional covariates is accounted for. This approach thus provides a principled way of disentangling genuine conditional links from spurious correlations driven by shared confounders.

\subsection{Quantile-on-Quantile Regression}
Quantile-on-Quantile Regression (QQR) enables us to explore how the dependence between happiness and the SDG Index varies across their entire joint distributions, while conditioning on other socioeconomic predictors. Unlike mean-based regressions, QQR estimates the effect of different quantiles of sustainability on different quantiles of happiness, thereby uncovering heterogeneous patterns that may be masked in the average relationship. This allows us to test, for example, whether sustainability improvements matter more for countries in the lower end of the happiness distribution, or whether the link strengthens among high-performing nations, providing a distributionally nuanced view of the happiness–sustainability nexus beyond the role of common covariates.

\subsubsection{QQR Model}
Let $\{(Y_i,X_i,Z_i)\}_{i=1}^n$ be i.i.d.\ observations, where $Y$ denotes (country-level) 
happiness, $X$ the SDG Index, and $Z \in \mathbb{R}^q$ a vector of controls 
(e.g., income,   life expectancy, governance). Denote by $F_X$ the distribution 
function of $X$ and by
\[
Q_X(\theta) \;=\; \inf\{x \in \mathbb{R}:\; F_X(x)\ge \theta\},\qquad \theta\in(0,1),
\]
the $\theta$-quantile of $X$. For $\tau \in (0,1)$, let $Q_{Y|X,Z}(\tau \mid x,z)$ be the 
conditional $\tau$-quantile of $Y$ given $(X,Z)=(x,z)$.

The QQR approach, introduced by \citet{sim2015quantile}, generalizes classical quantile 
regression \citep{koenker2005quantile} by studying how the $\tau$-quantile of $Y$ responds 
\emph{locally} to the behavior of $X$ around its $\theta$-quantile, producing a 
two-dimensional effect surface $(\tau,\theta)\mapsto \beta(\tau,\theta)$. Specifically, 
for $x$ near $x_\theta := Q_X(\theta)$, the conditional quantile function is approximated 
by the local linear expansion
\[
Q_{Y|X,Z}(\tau \mid x,z)
\;\approx\;
\alpha(\tau,\theta) \;+\; \beta(\tau,\theta)\,\big(x - x_\theta\big) \;+\; z^\top \gamma(\tau,\theta),
\]
where $\alpha(\tau,\theta)$ is a local intercept, $\beta(\tau,\theta)$ is the local slope 
(the parameter of interest), and $\gamma(\tau,\theta)$ captures the effect of controls.  

Given a kernel $K:\mathbb{R}\to \mathbb{R}_+$ and bandwidth $h>0$, define 
$K_h(u)=K(u/h)/h$ and the quantile loss $\rho_\tau(u) = u\,(\tau - \1\{u<0\})$, where $\1_A$ is the indicator function that takes 1 if condition $A$ is satisfied and zero otherwise. 
For each $(\tau,\theta)$, the QQR estimator 
$(\widehat{\alpha},\widehat{\beta},\widehat{\gamma})(\tau,\theta)$ solves the local 
weighted quantile regression
\[
(\widehat{\alpha},\widehat{\beta},\widehat{\gamma})(\tau,\theta)
\;=\;
\arg\min_{\alpha,\beta,\gamma}
\sum_{i=1}^n
\rho_\tau\!\left(
Y_i \;-\; \alpha \;-\; \beta\,(X_i - x_\theta) \;-\; Z_i^\top \gamma
\right)
\,K_h\!\big(X_i - x_\theta\big).
\]
The QQR \emph{effect surface} is then given by $\widehat{\beta}(\tau,\theta)$ on a grid 
$(\tau,\theta)\in\mathcal{T}\times\Theta \subset (0,1)^2$. By construction, 
$\widehat{\beta}(\tau,\theta)$ estimates the local partial derivative of the conditional 
$\tau$-quantile of $Y$ with respect to $X$ evaluated at $x_\theta$, i.e.,
\[
\beta(\tau,\theta) \;\approx\; \left.\frac{\partial}{\partial x}\, 
Q_{Y|X,Z}(\tau \mid x,z)\right|_{x=x_\theta}.
\]

Under standard smoothness and regularity conditions on $Q_{Y|X,Z}(\tau\mid x,z)$, the 
kernel $K$, and the bandwidth sequence $h=h_n\downarrow 0$ with $nh\to\infty$, the estimator 
$\widehat{\beta}(\tau,\theta)$ is consistent and asymptotically normal for fixed 
$(\tau,\theta)$ \citep{sim2015quantile}. Inference (e.g., confidence intervals or uniform 
bands) can be obtained via resampling methods tailored to local quantile regression 
\citep{chen2003inference}.  

QQR thus delivers a distributionally rich characterization of the dependence between 
happiness and sustainability: for fixed $\theta$ (countries around the $\theta$-quantile 
of SDG performance), $\widehat{\beta}(\tau,\theta)$ traces how SDG relates to different 
quantiles $\tau$ of happiness; varying $\theta$ reveals how this relationship changes 
across SDG regimes. Conditioning on $Z$ ensures that the estimated surface reflects the 
association \emph{beyond} common socioeconomic and institutional predictors, capturing 
potential nonlinearities and heterogeneous effects across the joint distribution of 
happiness and sustainability.

\section{Empirical Results}\label{res}

\subsection{Conditional dependence structure: Evidence from Graphical Lasso}

\begin{figure}[h!]
    \centering
    \includegraphics[width=0.5\textwidth]{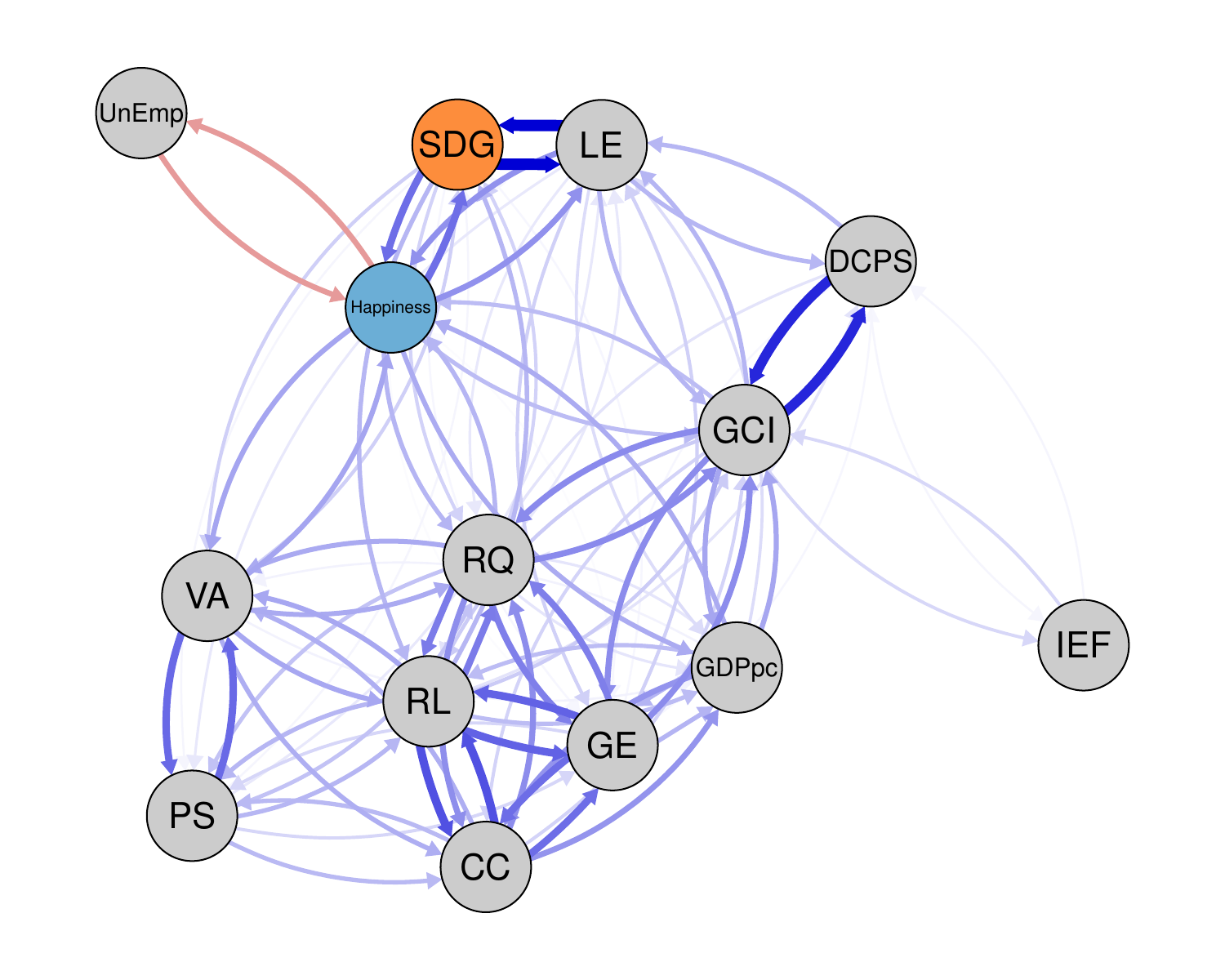}
    \caption{Graphical lasso conditional dependence network. Nodes represent variables, and edges indicate non-zero partial correlations after controlling for the rest of the system. Edge thickness reflects the magnitude of the partial correlation. Blue and orange nodes highlight Happiness and SDG, respectively.}
    \label{fig:glasso_network}
\end{figure}

The graphical lasso was applied to the complete-case dataset consisting of $p = 14$ variables and $n = 78$ country-level observations. This method estimates a sparse precision matrix, allowing us to identify direct conditional dependencies between variables while controlling for the influence of all other predictors. The results indicate that the edge between  {Happiness} and the  {SDG Index} is  {present}, suggesting a direct conditional link between the two variables even after adjusting for governance,   macroeconomic, and demographic covariates. The estimated partial correlation between happiness and the SDG Index is $\widehat{\rho}_{\text{Happiness,SDG}|\mathbf{X}} = 0.2082$, which is modest in magnitude but clearly distinct from zero.  \\
Figure~\ref{fig:glasso_network} displays the conditional dependence network obtained from the graphical lasso estimation based on the complete-case sample. Each node corresponds to a variable, and edges represent direct partial correlations after conditioning on all other variables. Blue edges represent positive conditional dependence, whereas orange edges indicate negative conditional dependence. The graph highlights several important patterns. First, {Happiness} and the  {SDG Index} are directly linked, confirming the presence of a conditional association even after controlling for governance, 
macroeconomic, and demographic covariates. This complements the numerical result showing a partial correlation of approximately $0.21$. Second, governance indicators-including control of corruption (CC), government effectiveness (GE), regulatory quality (RQ), rule of law (RL), and voice and accountability (VA)-form a densely connected cluster, reflecting their strong interdependence. Third,  {Life Expectancy (LE)} 
and  {GDP per capita (GDPpc)} emerge as central nodes, each connecting both to institutional quality and to outcome variables such as happiness and sustainability. \\ 
Overall, the network structure suggests that the happiness–sustainability nexus is embedded in a broader system of institutional and socioeconomic linkages. While governance variables are highly collinear and tightly clustered, happiness and sustainability retain a direct conditional link that is not fully explained by these shared predictors. This finding underscores the importance of analyzing dependence 
beyond pairwise correlations and provides further justification for exploring alternative methods such as quantile-on-quantile regression, which can capture nonlinearities and tail-specific dependencies.  

\subsection{Distributional Dependence: Evidence from QQR}
\begin{figure}[h!]
  \centering
  \begin{subfigure}[t]{0.49\textwidth}
    \centering
    \includegraphics[width=\linewidth]{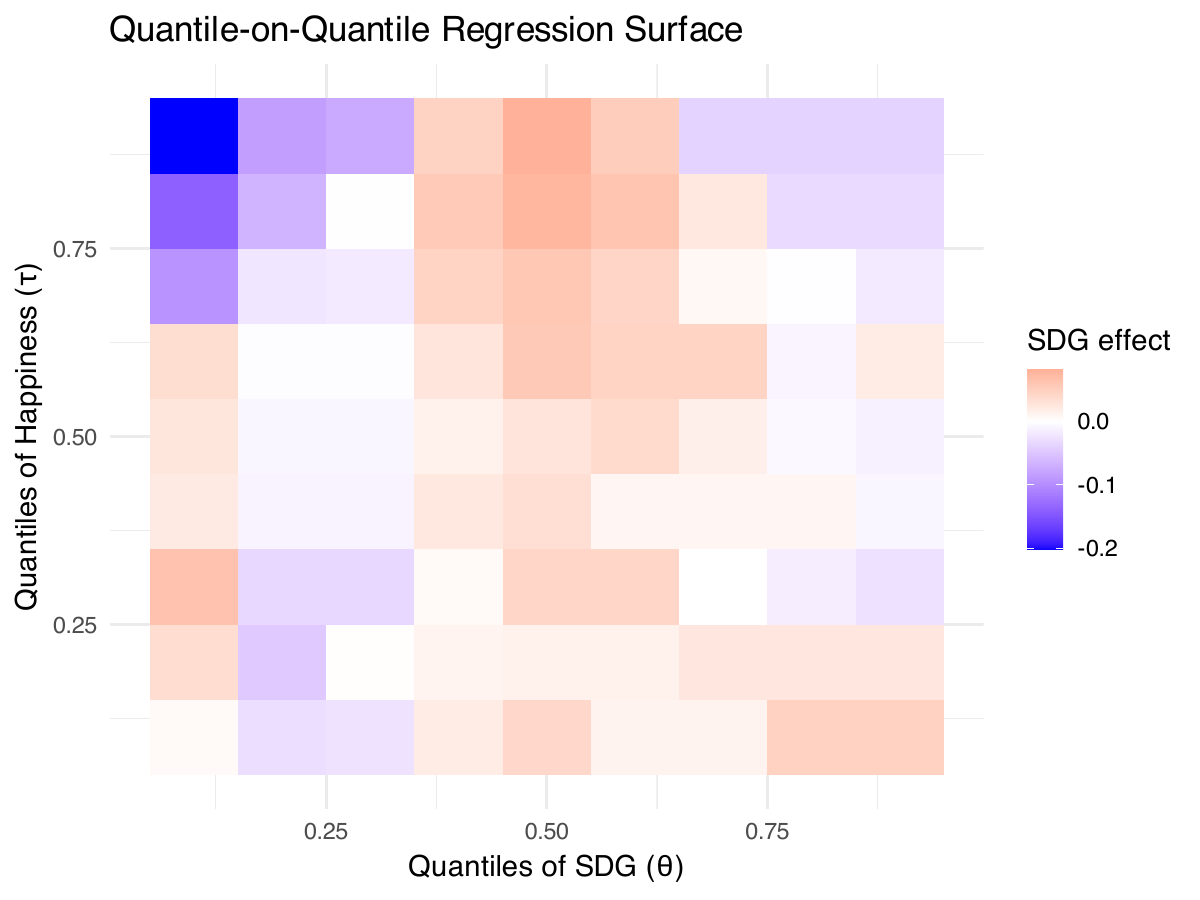}
    \caption{}
    \label{fig:qqr-heat_0109}
  \end{subfigure}
  \hfill
  \begin{subfigure}[t]{0.49\textwidth}
    \centering
    \includegraphics[width=\linewidth, height = 8cm]{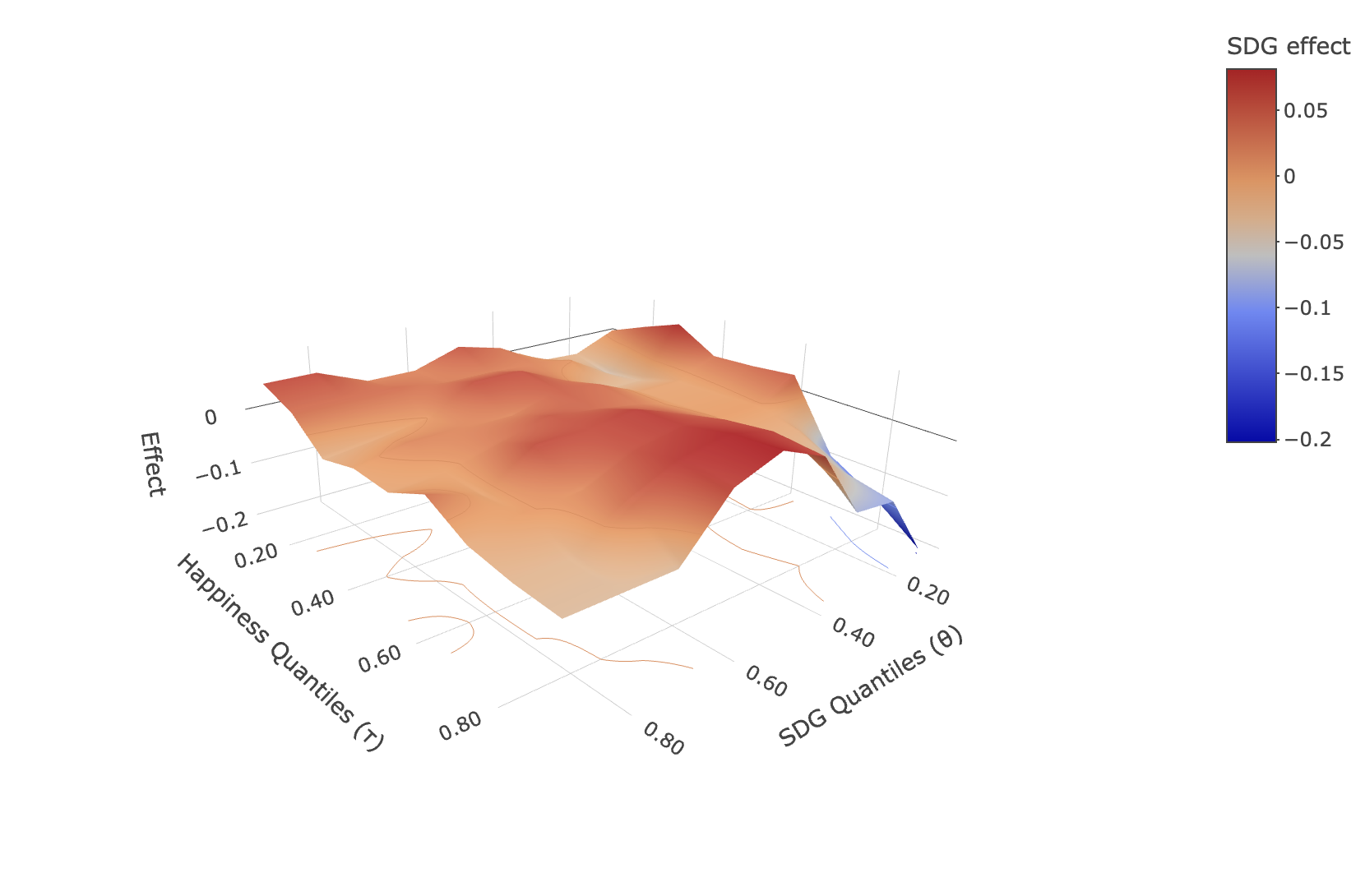}
    \caption{}
    \label{fig:qqr-3d_0109}
  \end{subfigure}
  \caption{Quantile-on-Quantile Regression results shown as (a) a 2D heatmap and
  (b) a 3D surface. Warmer colors denote more positive SDG effects on happiness; cooler colors denote negative effects. Here $(\tau, \theta) \in [0.1, 0.9].$}
  \label{fig:qqr_surface_01_09}
\end{figure}

In order to implement the quantile-on-quantile regression, two alternative grids of quantiles were considered: a fine grid ranging from $0.01$ to $0.99$ and a more central grid ranging from $0.10$ to $0.90$. The first specification captures the extreme tails of both the happiness and sustainability distributions, thereby highlighting the behavior of outlier countries such as resource-rich economies with relatively high happiness but very low sustainability scores.\\
Figure~\ref{fig:qqr_surface_01_09} displays the quantile-on-quantile regression surface estimated over the $0.1$–$0.9$ grid of happiness and SDG quantiles. The analysis of the surface reveals substantial heterogeneity across the distribution. In the  {lower-left corner} (low happiness, low sustainability), the effect of SDG on happiness is weakly negative, indicating that in contexts where both dimensions are at the bottom of their distributions, marginal gains in sustainability do not translate into improved well-being. In the  {lower-right corner} (low happiness, high sustainability), the effect becomes modestly positive, suggesting that for 
countries with relatively strong sustainability performance but low reported well-being, further progress on SDGs is associated with improvements in happiness. In the  {upper-left corner} (high happiness, low sustainability), the effect turns sharply negative, as highlighted by the dark blue region of the surface, implying that in countries where happiness is relatively high despite weak sustainability, increases in SDG scores may coincide with declines in happiness, perhaps reflecting transitional trade-offs or structural imbalances. In contrast, the  {upper-right corner} (high happiness, high sustainability) shows effects close to zero, indicating that once both dimensions are simultaneously strong, marginal changes in sustainability 
do not meaningfully alter happiness. Finally, in the  {center of the distribution} 
(around $\tau = 0.5$, $\theta = 0.5$), the surface is nearly flat, reflecting a neutral relationship at typical levels of happiness and sustainability. Taken together, these results emphasize that the happiness–sustainability nexus is driven by asymmetric tail effects: strongly negative in the upper-left corner, modestly positive in the lower-right corner, and essentially neutral elsewhere.  

\begin{figure}[h!]
  \centering
  \begin{subfigure}[t]{0.49\textwidth}
    \centering
    \includegraphics[width=\linewidth]{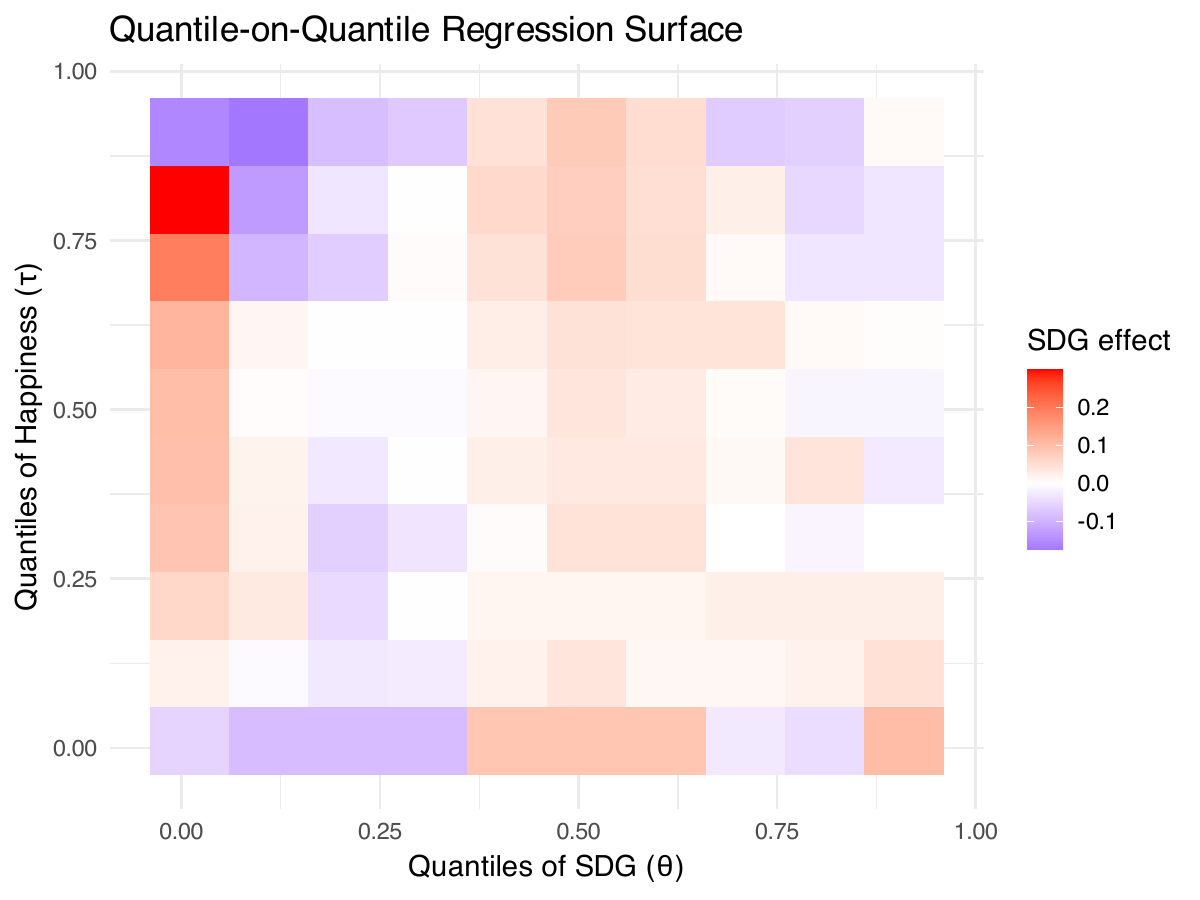}
    \caption{}
    \label{fig:qqr-heat}
  \end{subfigure}
  \hfill
  \begin{subfigure}[t]{0.49\textwidth}
    \centering
    \includegraphics[width=\linewidth, height = 7cm]{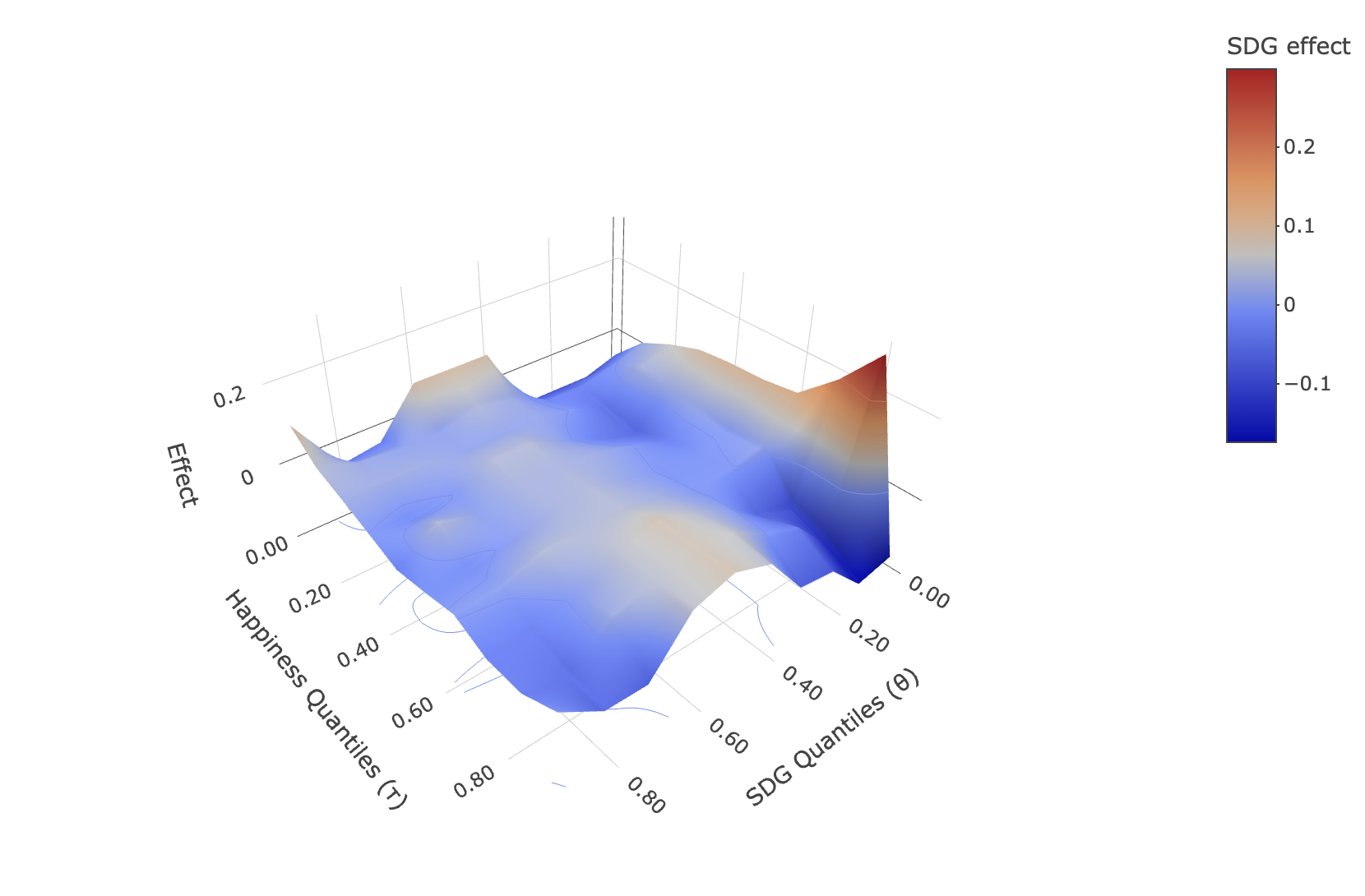}
    \caption{}
    \label{fig:qqr-3d}
  \end{subfigure}
  \caption{Quantile-on-Quantile Regression results shown as (a) a 2D heatmap and
  (b) a 3D surface. Here $(\tau, \theta) \in [0.01, 0.99].$}
  \label{fig:qqr_surface_001_099}
\end{figure}

Figure~\ref{fig:qqr_surface_001_099} reports the QQR surface estimated over the $0.01$–$0.99$ grid, thereby capturing the extreme tails of both distributions. The surface shows clear heterogeneity across regions of the joint distribution. In the  {lower-left corner} (low happiness, low sustainability), the effect of SDG on happiness is slightly negative, suggesting that for countries where both dimensions are weak, further gains in sustainability do not directly translate into improvements in well-being. In the  {lower-right corner} (low happiness, high sustainability), the effect becomes moderately positive, implying that countries with solid sustainability performance but low happiness levels benefit in terms of well-being from additional SDG progress. In the  {upper-left corner} (high happiness, low sustainability), the effect is sharply negative, consistent with the idea that countries achieving high happiness in spite of poor sustainability may face structural trade-offs, whereby improvements in sustainability indicators are associated with reductions 
in perceived well-being. Conversely, the  {upper-right corner} (high happiness, high sustainability) shows effects close to zero, indicating that once both dimensions are simultaneously strong, marginal changes in sustainability have little additional impact on happiness. Finally, in the  {center of the distribution} (around $\tau = 0.5$, $\theta = 0.5$), the surface is essentially flat, suggesting a neutral relationship at moderate levels of both variables. \\
In the first column of the QQR surface in Figure \ref{fig:qqr_surface_001_099}, corresponding to very low sustainability quantiles ($\theta < 0.1$), the effect of SDG on happiness is predominantly positive (red shading) for moderate-to-high happiness quantiles ($\tau \in [0.25,0.90]$). This indicates that in countries with chronically weak sustainability performance, marginal improvements in SDG outcomes are associated with increases in happiness when the population already reports moderate to high levels of well-being. From an economic perspective, this finding suggests that even small sustainability gains-such as progress in healthcare, education, or infrastructure-reinforce and complement well-being in societies where happiness is relatively secure. The result highlights an asymmetric feature of the nexus: while very high happiness in extremely unsustainable contexts may generate negative trade-offs, moderate-to-high happiness levels appear to benefit positively from incremental sustainability improvements when starting from a low baseline. 

\section{Conclusion and Policy implications}\label{conc}

This paper revisited the relationship between happiness and sustainability by leveraging two complementary approaches—Graphical Lasso, and Quantile-on-Quantile Regression—applied to recent cross-country data. Our results converge on three main insights. First, the conditional network analysis using Graphical Lasso demonstrates that happiness and the SDG Index share a direct link after accounting for key socioeconomic and governance covariates, though the partial correlation is modest in magnitude. Then, the QQR analysis reveals substantial heterogeneity across the joint distribution: sustainability improvements appear to enhance happiness in low-happiness, high-sustainability countries, reduce it in high-happiness, low-sustainability contexts, and leave it largely unchanged in the central mass of the distribution.\\
Taken together, these findings imply that the happiness–sustainability nexus cannot be captured by a single linear or mean-based model. Instead, the relationship is modest in magnitude, shaped by common drivers such as income, governance, and life expectancy, and exhibits asymmetric effects at distributional extremes. The evidence therefore calls for a more nuanced policy perspective. Governments should not assume that improvements in sustainability will automatically translate into higher well-being across all contexts. 
Rather, the alignment between the two dimensions depends crucially on a country’s baseline position in the joint distribution of happiness and sustainability.\\
 \textbf{Policy implications.} First, for countries that score high on sustainability but lag behind in well-being, targeted social policies may be necessary to ensure that sustainability achievements translate into tangible improvements in people’s lives. This could include expanding social protection, addressing   or investing in human capital so that sustainability gains are perceived as directly enhancing life satisfaction. Second, for countries with high levels of happiness but low sustainability performance, policymakers must anticipate potential trade-offs: transitioning to more sustainable practices may initially reduce satisfaction if it requires lifestyle adjustments or short-term economic sacrifices. Communication, public engagement, and compensation mechanisms are therefore essential to manage these transitions without 
eroding trust. Third, for the majority of countries clustered around the middle of the distribution, where the relationship is essentially neutral, the focus should be on strengthening institutional quality, governance, and income growth, since these remain the strongest common drivers of both happiness and sustainability. Finally, international organizations should recognize that a ``one-size-fits-all" model is inappropriate: global agendas such as the SDGs must be adapted to national circumstances, with a 
clear understanding of where synergies with well-being exist and where tensions may arise.\\
Overall, our analysis underscores that the happiness–sustainability nexus is complex, modest, and context-dependent. Policies aiming to simultaneously improve well-being and sustainability will only succeed if they acknowledge this heterogeneity and place institutional and socioeconomic drivers at the core of their design. By doing so, policymakers can move beyond simplistic expectations and build strategies that are both sustainable and capable of enhancing human flourishing.\\




\bibliographystyle{apalike}
\bibliography{sample}

\section*{Appendix: Variables and Data Sources}
The empirical analysis relies on a set of economic, social, and governance indicators drawn from widely used international databases. The two outcome variables of primary interest are \textit{Happiness} and the \textit{SDG Index}.  

\begin{enumerate}
    \item \textbf{Variable: Happiness} \\
    \textbf{Code:} Happiness \\
    \textbf{Website:} World Happiness Report \\
    \textbf{Definition:} Self-reported Cantril ladder scale (0--10), where 0 denotes the worst possible life and 10 the best possible life.  

    \item \textbf{Variable: SDG Index} \\
    \textbf{Code:} SDG \\
    \textbf{Website:} \texttt{sdgindex.org} \\
    \textbf{Definition:} Composite measure of progress across the 17 Sustainable Development Goals.  
\end{enumerate}

To account for economic and demographic influences:  

\begin{enumerate}
    \setcounter{enumi}{2}
    \item \textbf{Variable: GDP per capita (GDPpc)} \\
    \textbf{Code:} GDPpc \\
    \textbf{Website:} World Bank WDI \\
    \textbf{Transformation:} constant 2015 US dollars, logarithmic.  

    \item \textbf{Variable: Domestic Credit to the Private Sector (DCPS)} \\
    \textbf{Code:} DCPS \\
    \textbf{Website:} World Bank WDI \\
    \textbf{Definition:} Financial resources provided to the private sector by financial corporations.  

    \item \textbf{Variable: Life Expectancy (LE)} \\
    \textbf{Code:} LE \\
    \textbf{Website:} World Bank WDI \\
    \textbf{Definition:} Average number of years a newborn infant would live if prevailing mortality patterns persist.  

    \item \textbf{Variable: Unemployment Rate} \\
    \textbf{Code:} UnEmp \\
    \textbf{Website:} World Bank WDI \\
    \textbf{Definition:} Share of the labor force without work but available for and seeking employment.  

    \item \textbf{Variable: Index of Economic Freedom (IEF)} \\
    \textbf{Code:} IEF \\
    \textbf{Website:} \texttt{heritage.org} \\
    \textbf{Definition:} Composite index covering property rights, fiscal health, government integrity, business and trade freedom.  
\end{enumerate}

For institutional quality and governance, we rely on the six Worldwide Governance Indicators (WGI):  

\begin{enumerate}
    \setcounter{enumi}{7}
    \item \textbf{Variable: Control of Corruption (CC)} \\
    \textbf{Code:} CC \\
    \textbf{Website:} World Bank WGI \\
    \textbf{Definition:} Perceptions of the extent to which public power is exercised for private gain.  

    \item \textbf{Variable: Government Effectiveness (GE)} \\
    \textbf{Code:} GE \\
    \textbf{Website:} World Bank WGI \\
    \textbf{Definition:} Quality of public services, policy formulation, and credibility of government commitments.  

    \item \textbf{Variable: Political Stability and Absence of Violence/Terrorism} \\
    \textbf{Code:} PS \\
    \textbf{Website:} World Bank WGI \\
    \textbf{Definition:} Likelihood of political instability and politically motivated violence.  

    \item \textbf{Variable: Regulatory Quality} \\
    \textbf{Code:} RQ \\
    \textbf{Website:} World Bank WGI \\
    \textbf{Definition:} Government’s ability to formulate and implement sound policies and regulations.  

    \item \textbf{Variable: Rule of Law} \\
    \textbf{Code:} RL \\
    \textbf{Website:} World Bank WGI \\
    \textbf{Definition:} Confidence in legal institutions, contract enforcement, and property rights.  

    \item \textbf{Variable: Voice and Accountability} \\
    \textbf{Code:} VA \\
    \textbf{Website:} World Bank WGI \\
    \textbf{Definition:} Extent to which citizens participate in selecting governments and enjoy freedoms of expression, association, and media.  
\end{enumerate}

Finally, competitiveness:  

\begin{enumerate}
    \setcounter{enumi}{13}
    \item \textbf{Variable: Global Competitiveness Index} \\
    \textbf{Code:} GCI \\
    \textbf{Website:} World Bank Group \\
    \textbf{Definition:} Summarizes institutions, policies, and factors determining national productivity.  
\end{enumerate}


\end{document}